\documentclass[a4paper]{article}%
\usepackage{amsmath}
\usepackage{mathrsfs}
\usepackage{array}
\usepackage{lipsum,appendix}
\usepackage{amsfonts}
\usepackage{amssymb}
\usepackage[lmargin=2.0cm,rmargin=3cm,tmargin=3.0cm,bmargin=2.0cm]{geometry} 
\usepackage{graphicx}%
\usepackage{booktabs}

\begin{document}

\title{Combined effect of NSI and SFP on solar electron neutrino oscillation} 

\author{Deniz Yilmaz\\ \textit{Department of Physics Engineering, Faculty of Engineering, Ankara University}\\ \textit{06100 Tandogan, Ankara, TURKEY}\\ \textit{e-mail:dyilmaz@eng.ankara.edu.tr} }
\providecommand{\keywords}[1]{\textbf{\text{Key words:}} #1}
\maketitle

\begin{abstract}
The combined effect of Spin Flavor Precession (SFP) and the non standard neutrino interaction (NSI) on the survival probability of solar electron neutrinos (assumed to be Dirac particles) is examined for 
various values of $\epsilon_{11}$, $\epsilon_{12}$ and $\mu B$. It is found that the neutrino survival probability curves affected by SFP and NSI effects individually for some values of the parameters 
($\epsilon_{11}$, $\epsilon_{12}$ and $\mu B$) get close to the standard MSW curve when both effects are combined. Therefore, the combined effect of SFP and NSI needs to be taken into account when the solar electron neutrino 
data obtained by low energy solar neutrino experiments is investigated.
\end{abstract}


\section{INTRODUCTION}
After first observation of the solar neutrino oscillation in Homestake neutrino experiment, serious solar, atmospheric and reactor neutrino 
experiments were established to confirm it during the last decades. Both KamLAND experiment detecting reactor neutrinos [1, 2] and 
the combined analysis of the solar neutrino experiments (high precision water Cherenkov experiments SNO [3, 4] and SK [5, 6] and the 
radiochemical experiments Homestake [7], SAGE [8], GALLEX [9] and GNO [10]) strongly pointed out the so-called large mixing angle (LMA) region of the neutrino parameter 
space [11-16]. One of the implications of the physics beyond the Standard Model (SM) is the neutrino oscillation. Since neutrinos have a mass 
in a minimal extension of the SM, they have also magnetic moment [17]:
\begin{equation}
\mu_{\nu}=\frac{3eG_{f}m_{\nu}}{8\pi^{2}\sqrt{2}}=\frac{3eG_{f}m_{e}m_{\nu}}{4\pi^{2}\sqrt{2}}\mu_{B}
\end{equation}
where $G_f$ is Fermi constant; $m_e$ and $m_\nu$ are the masses of electron and neutrino, respectively; and $\mu_{B}$ is Bohr magneton.
While Majorana type neutrinos can only have off-diagonal (transition) magnetic moments, Dirac type neutrinos can have diagonal and off diagonal magnetic moments [18, 19]. 
If the neutrinos have magnetic moments they can be effected by the large magnetic fields when they are 
passing through the magnetic region. Their spin can flip and the left-handed neutrino becomes a right-handed neutrino [20-24]. 
Thus the combined effect of the matter and the magnetic field called as spin flavor precession (SFP) can change left-handed electron 
neutrino to the right-handed another neutrino. This yields two other transitions ($\nu_{e_L}\rightarrow\nu_{\mu_R} \text{ or } \nu_{\tau_R} $) in addition to the left-handed ones 
( i.e. in this scenario, the conversion probability is mainly affected) [24]. In the Dirac case, since the right-handed neutrinos are considered as sterile, they are not detectable by the detectors. On the other hand, if the neutrinos are of Majorana type, this conversion yields a solar antineutrino flux which are 
detectable by the detectors. These conversions for both Dirac and Majorana cases can also be responsible for the solar electron neutrino deficit. 
So far several studies related with the SFP have been studied in different aspects [25-31].
Astrophysical and cosmological arguments [32], Supernova 1987A [33, 34], solar neutrino experiments looking neutrino-electron scattering [35] and
the reactor neutrino experiments [36, 37] provide some bounds on the neutrino magnetic moment. 
The new limit recently was obtained by GEMMA experiment: $\mu_{\nu}<2.9\times10^{-11}\mu_{B}$ at $90\%$ CL [38]. 
However, another strong bound on neutrino transition magnetic moment was obtained in the presence of non-standard neutrino-nucleus interactions by Papoulias and Kosmas [39]. 
Detailed discussion on neutrino magnetic moment is also given elsewhere [40-45]. 
In additon to the konowledge about neutrino magnetic moment, the thorough information of solar magnetic fields is needed for the SFP analysis in the Sun.
Even though the limited knowledge about it, some plausible profiles can be found in the literature [46, 47]. 
Standard solar model [47, 48] limits the solar magnetic field: $\sim$ 20 G near the solar surface [49], 20 kG - 300 kG at the convective zone [47] 
and $<  10^7 $ G at the solar center [47]. In this study the magnetic field profile is choosen as given in Ref. [46]. 
It has a peak at the bottom of the convective zone as shown in figure 1. 

Solar neutrinos can also be used for analysing of the physics beyond the Standard Model of the particle physics such as 
non-standard forward scattering [50], mass varrying neutrinos [51, 52] and long-range leptonic forces [53]. 
The probe of non-standard neutrino interaction models is expected to observe in the transition region between 1 MeV and 4 MeV where 
the low energy solar neutrino experiments such as SNO+ will examine. Even though the data is poor in this region, the studies 
comparing the effects of non-standard models on the neutrino oscillation to the standard MSW-LMA oscillation shows that these effects 
modify the survival probability of neutrinos [50-56].  

In this letter, the combined effect of non-standard neutrino interactions (NSI) and SFP 
is examined in the case of two neutrino generations by assuming that the neutrinos are of Dirac type. 
The best fit LMA values are used for $\delta m^2_{12}$ and $\theta_{12}$ [57]. 
It is shown that the neutrino survival probability curves affected by SFP and NSI effects individually for some values of the parameters 
($\epsilon_{11}$, $\epsilon_{12}$ and $\mu B$) get close to the standard MSW curve when both effects are combined.
Therefore, one can say that the combined effect of them needs to be taken into account  when the solar electron neutrino 
data obtained by low energy solar neutrino experiments is investigated. 
Another analysis on the SFP effect in the presence of the NSI is examined for Majorana type solar neutrinos in Ref. [58]. 

\begin{figure}
[t]
\begin{center}
\includegraphics[height=6cm,width=6cm]{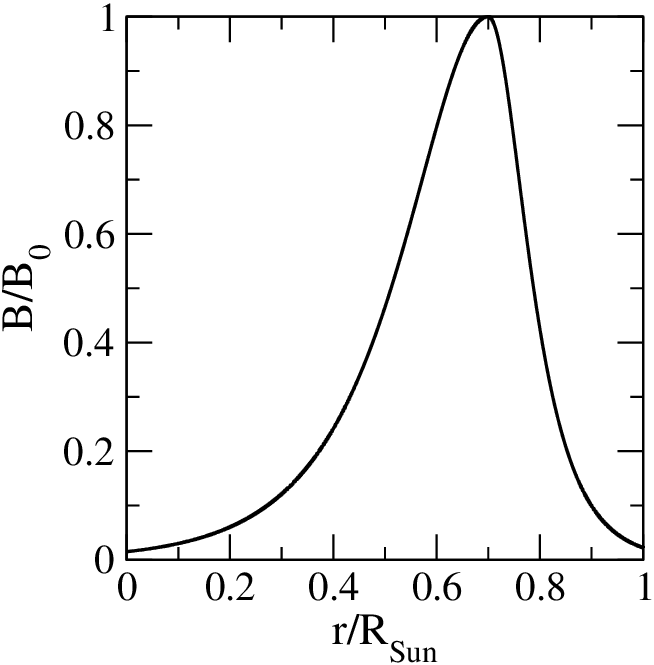}%
\caption{Magnetic field profile.} \label{fig:fig1}
\end{center}
\end{figure}
\section{Spin Flavor Precession (SFP) Including Non-Standard Neutrino Interaction (NSI)}
The evolution equation including NSI matter effects in the SFP scenario for Dirac neutrinos can be written as 
\begin{equation}
i\frac{d}{dt}\left(
\begin{array}
[c]{c}%
\nu _{e_L}\\
\nu_{\mu_L}\\
\nu_{e_R}\\
\nu_{\mu_R}\\
\end{array}
\right) =\left(
\begin{array}{cc}
 H_L+H_{NSI} & BM^\dagger\\
BM & H_R  \\
\end{array}
\right)\left(
\begin{array}
[c]{c}%
\nu _{e_L}\\
\nu_{\mu_L}\\
\nu_{e_R}\\
\nu_{\mu_R}\\
\end{array}
\right), 
\end{equation}
here $H_L$, $H_R$, $H_{NSI}$ and $M$ are the $2\times2$ submatrices and $B$ is the transverse magnetic field [24, 50]. For the Dirac neutrinos one writes down,
\begin{equation}
H_L=\left(
\begin{array}{cc}
V_{c}+V_{n}+\dfrac{\delta m^{2}_{12}}{2E} sin^2\theta_{12}  & \dfrac{\delta m^{2}_{12}}{4E} sin 2\theta_{12}  \\
\dfrac{\delta m^{2}_{12}}{4E} sin 2\theta_{12}  & V_{n}+\dfrac{\delta m^{2}_{12}}{2E} cos^2\theta_{12}  \\
\end{array}
\right),
\end{equation}
and $H_R=H_L (V_c=0=V_n)$. The matter potentials here are given as
\begin{equation}
V_c=\sqrt{2}G_F N_e,\quad V_n=-\frac{G_F}{\sqrt{2}} N_n,
\end{equation}
where $N_e$ and $N_n$ are electron and neutron density, respectively [59-61]. 
The magnetic moment matrix for the Dirac neutrinos in the Eq. (2) is written as [24] 
\begin{equation}
M=\left( \begin{array}
[c]{cc}%
\mu_{ee}  & \mu_{e\mu}  \\
\mu_{\mu e}  & \mu_{\mu \mu}  \\
\end{array}\right).
\end{equation}
The NSI contributions in Eq. (2) can be parametrized by four-fermion operator as given in Ref. [50]
\begin{equation}
\mathcal{L}=-2\sqrt{2}G_F(\nu_\alpha\gamma_\rho\nu_\beta)(\epsilon^{f\bar{f}L}_{\alpha\beta}\bar{f}_L\gamma^\rho\bar{f}_L+\epsilon^{f\bar{f}R}_{\alpha\beta}\bar{f}_R\gamma^\rho\bar{f}_R)
,\end{equation}
here $\epsilon^{f\bar{f}P}$ denotes the strength of the non-standard interaction between the $\alpha$ and $\beta$ types of neutrinos and the P(left or right)-handed components of the fermions $f$ and $\bar{f}$. 
Since the neutrino propagation can only be effected by the vector components where $f=\bar{f}$ of the non-standard interaction ($\epsilon^{f}_{\alpha\beta}=\epsilon^{ffL}_{\alpha\beta}+\epsilon^{ffR}_{\alpha\beta}$), one can define the $\epsilon_{\alpha\beta}$ as the sum of the contributions from electrons, up quarks and down quarks in matter: $\epsilon_{\alpha\beta}=\sum_{f=e,u,d}\epsilon^{f}_{\alpha\beta}N_f/N_e$. 
Then, the three flavor NSI Hamiltonian can be written as
\begin{equation}
 H^{3\times3}_{NSI}=V_c\left( \begin{array}
[c]{ccc}%
\epsilon_{ee}  & \epsilon^\ast_{e\mu} & \epsilon^\ast_{e \tau} \\
\epsilon_{e \mu}  & \epsilon_{\mu \mu} & \epsilon^\ast_{\mu \tau}  \\
\epsilon_{e \tau} & \epsilon_{\mu \tau} & \epsilon_{\tau \tau} \\
\end{array}\right).
\end{equation}
After performing a rotation to $H^{3\times3}_{NSI}$ by using the two factor of the neutrino mixing matrix, $T_{13}T_{23}$, 
\begin{equation}
T^\dagger_{13}T^\dagger_{23}H^{3\times3}_{NSI}T_{13}T_{23}
\end{equation}
and decoupling the third flavor as in the standard three flavor neutrino oscillation calculations, one can find the $2\times2$ neutrino non-standard interaction (NSI) part in Eq. (2) as 
\begin{equation}
H_{NSI}=V_{c}\left(
\begin{array}{cc}
0 & \epsilon^*_{12} \\
\epsilon_{12} &  \epsilon_{11}\\
\end{array}
\right)
\end{equation}
where $\epsilon_{11}$ and $\epsilon_{12}$ are the contributions from the new physics related to the original vectorial couplings,
$\epsilon_{\alpha\beta}$, given as 
\begin{equation}
\begin{aligned}
\epsilon_{11}={} & \ \epsilon_{\mu\mu}c^2_{23}-(\epsilon_{\mu\tau}+\epsilon^{\ast}_{\mu\tau})s_{23}c_{23}+\epsilon_{\tau\tau}s^2_{23}-\epsilon_{ee}c^2_{13}+s_{13}[(e^{-i\delta}\epsilon_{e\mu}+e^{i\delta}\epsilon^{\ast}_{e\mu})c_{13}s_{23}\\
                 &  +(e^{-i\delta}\epsilon_{e\tau}+e^{i\delta}\epsilon^{\ast}_{e\tau})c_{13}c_{23}]-s^2_{13}[(\epsilon_{\mu\tau}+\epsilon^{\ast}_{\mu\tau})s_{23}c_{23}+\epsilon_{\mu\mu}s^2_{23}+\epsilon_{\tau\tau}c^2_{23}], 
\end{aligned}
\end{equation}
and 
\begin{equation}
\epsilon_{12}=c_{13}(\epsilon_{e\mu}c_{23}-\epsilon_{e\tau}s_{23})+s_{13}e^{i\delta}[\epsilon_{\mu\tau}s^2_{23}-\epsilon^{\ast}_{\mu\tau}c^2_{23}-(\epsilon_{\mu\mu}-\epsilon_{\tau\tau})s_{23}c_{23}].
\end{equation}
Here $c_{ij}=cos\theta_{ij}$ and $s_{ij}=sin\theta_{ij}$ and the $\delta$ is the CP-violating phase that we will ignore in our discussion [54]. 

The direct bounds on the NSI parameters come from athmospheric neutrino experiments (Super Kamiokande, IceCube-79) [62, 63],
accelerator neutrino experiments (MINOS) [64] and some phenomenological studies [65-68]: 
$|\epsilon_{ee}|\lesssim0.5$ [62], $|\epsilon_{e\tau}|\lesssim0.5$ [62],  $|\epsilon_{\mu\tau}|\lesssim6\times10^{-3}$ [63], $|\epsilon_{\tau\tau}-\epsilon_{\mu\mu}|\lesssim3\times10^{-2}$ [63], 
$-0.067\lesssim\epsilon_{\mu\tau}\lesssim0.023$ [64].
The effect of NSI were also studied by using data of reactor neutrino experiment, DayaBAY, [69] and solar neutrino experiments [70].
Detailed analysis on the non-standard neutrino interactions and their limits is given in Ref. [71] and Ref. [72].

\section{Results and Conclusions}

\begin{figure}
[t]
\begin{center}
\includegraphics[height=11cm,width=14cm]{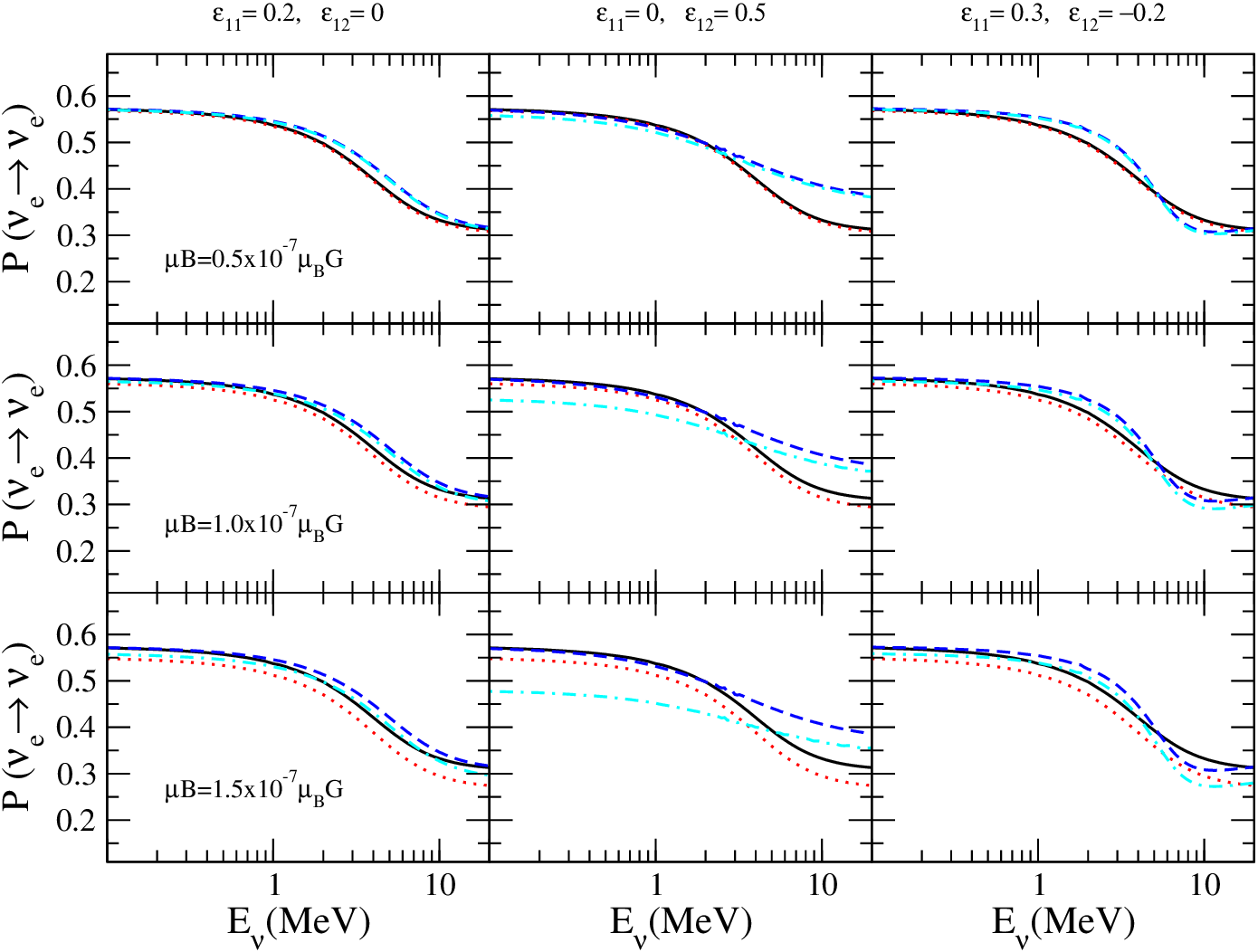}%
\caption{Survival probabilities for MSW-LMA prediction alone (solid lines), SFP effect at different $\mu B$ values (dotted lines), 
NSI effect alone (dashed lines) and the combined effect of the NSI and SFP (dotted-dashed lines).
Each column uses the same  $\epsilon_{11}$ and $\epsilon_{12}$ values, each row uses the same $\mu B$ values. } \label{fig:fig2}
\end{center}
\end{figure}
In this analysis the combined effect of the non-standard neutrino interaction and SFP on the survival probability of 
solar electron neutrinos (assumed to be Dirac particles) is examined for various values of $\epsilon_{11}$, $\epsilon_{12}$ and $\mu B$. 
Results presented here are obtained numerically by diagonalizing the Hamiltonian in equation (2).
In the calculations, the magnetic field profile given in figure 1 is choosen as a Gaussian shape extending over the entire Sun [46] and 
the MSW-LMA best fit values are used: $\delta m^2_{12}=7.54\times 10^{-5} eV^2$ and $sin^2\theta_{12}=0.308$ [57].

Electron neutrino survival probabilities plotted as a function of neutrino energy are shown in figure 2 for all situations: 
MSW-LMA prediction alone (solid lines), SFP alone (dotted lines), MSW-LMA + NSI (dashed lines) and SFP + NSI (dotted-dashed lines).  
In this figure, different from the SFP effect seen for all neutrino energies, the new physics effects changes the standard MSW-LMA curve especially at 
the energies of $E\gtrsim1$ MeV in which the region of $E\gtrsim3.5$ MeV is well examined by the solar neutrino experiments SNO and SK. 
When the combined effect of them (SFP + NSI) is considered, the curves get closer to the the standard curve than the curves affected by them individually for some values of the parameters
($\epsilon_{11}$, $\epsilon_{12}$ and $\mu B$). A similar result was found in the another analysis examined for Majorana neutrinos for only one NSI parameter, $\epsilon_{12}$ [58].
However, compared to the Dirac case presented here, SFP effect is seen at almost ten times larger $\mu B$ values in the Majorana case.

The allowed regions obtained by using the SNO results [73] are shown in figure 3 in the ($\varepsilon_{11}, \mu B$) and ($\varepsilon_{12}, \mu B$) planes at 90\% CL for 10 MeV neutrino energy.
Even though the values of NSI parameters are expected to be very small ($\lesssim10^{-2}$), 
the large values of them is in the allowed regions when considering the SFP and NSI effects together.
It is seen that the current solar neutrino data constrain the $\mu B$ and ($\varepsilon_{11},\varepsilon_{12}$) values poorly. 
A practical limit on them can be expected from the data obtained by the new low energy ($1 \text{MeV} \lesssim E \lesssim 4$ MeV) solar neutrino experiments such as SNO+ [74] probing the evidence of new physics effect.
However, as it can be seen from the analysis presented here, the combined effect of SFP and NSI needs to be taken into account when the solar electron neutrino data obtained by 
new solar neutrino experiments is analysed.

\begin{figure}
[t]
\begin{center}
\includegraphics[height=6cm,width=13cm]{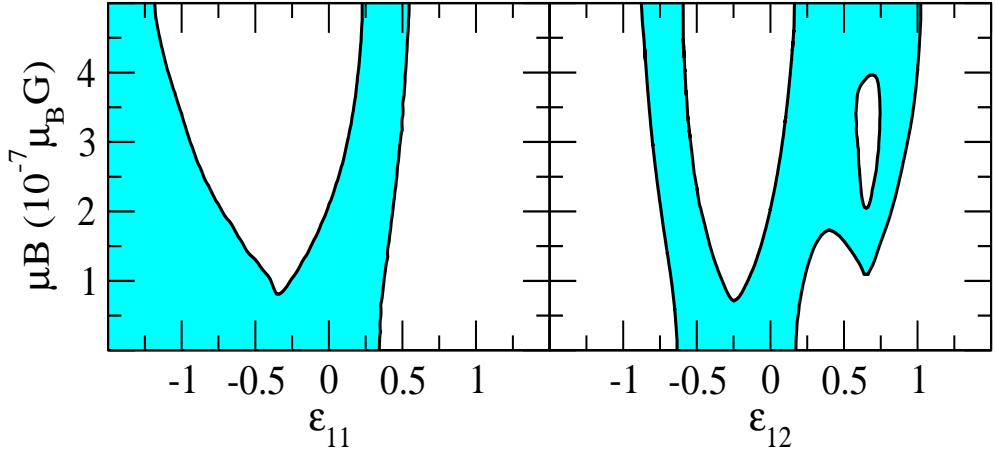}%
\caption{Allowed regions in the ($\varepsilon_{11}, \mu B$) and ($\varepsilon_{12}, \mu B$) planes at 90\% CL for 10 MeV neutrino energy. } \label{fig:fig3}
\end{center}
\end{figure}

\section*{Conflict of Interests}

The author declares that there is no conflict of interests regarding the publication of this paper.

\clearpage

\end{document}